\documentclass[journal]{vgtc}                     

\usepackage{graphicx}    
\usepackage{microtype}                 
\PassOptionsToPackage{warn}{textcomp}  
\usepackage{textcomp}                  
\usepackage{mathptmx}                  
\usepackage{times}                     
\usepackage{cite}                      
\usepackage{tabu}                      
\usepackage{booktabs}                  
\usepackage[table,xcdraw,svgnames]{xcolor}

\usepackage{multirow}
\usepackage{colortbl}
\usepackage{adjustbox}
\usepackage{tabularray}
\usepackage{adjustbox}
\usepackage{comment}
\usepackage{float}
\usepackage{array}
\usepackage{pgfplots}
\usepackage{tikz}
\usepackage{hhline}
\usepackage{arydshln}

\definecolor{cerulean}{rgb}{0.0,0.48,0.65}
\definecolor{caribbean}{rgb}{0.0,0.8,0.6}
\definecolor{paleblue}{rgb}{0.82,0.93,1}
\definecolor{auburn}{rgb}{0.43,0.21,0.1}
\definecolor{byzantium}{rgb}{0.44,0.16,0.39}
\definecolor{byzantine}{rgb}{0.74,0.2,0.64}
\definecolor{violet}{rgb}{0.54,0.17,0.89}
\definecolor{cornell}{rgb}{0.7,0.11,0.11}
\definecolor{lgray}{gray}{0.85}
\definecolor{mgray}{gray}{0.7}
\definecolor{dgreen}{rgb}{0,0.55,0}
\definecolor{mgreen}{rgb}{0,0.7,0}
\definecolor{palered}{rgb}{1,0.70,0.75}
\definecolor{pastelyellow}{rgb}{0.99,0.99,0.65}
\definecolor{purple}{rgb}{0.63,0.13,0.94}

\newcommand{\City}{{Brisbane}}
\newcommand{\State}{{Victoria}}
\newcommand{\Country}{{Australia}}

\newcommand{\eg}{e.g.,}
\newcommand{\ie}{i.e.,}
\newcommand{\etal}{{\em et al.}}

\newcommand{\Stock}{{\em Stock}}
\newcommand{\Water}{{\em Water}}
\newcommand{\Fire}{{\em Fire}}

\newcommand{\bipstart}[1]{\smallskip\noindent{\textbf{\textit{#1.}}}}
\newcommand{\myitem}{\vspace*{-1mm}\item}

\labelwidth\leftmargini\advance\labelwidth-\labelsep \labelsep 5pt
\def\@listi{\leftmargin\leftmargini
   \labelwidth\leftmarginii\advance\labelwidth-\labelsep
   \topsep 0pt plus 1pt minus 0.5pt
   \parsep 0pt plus 0.5pt minus 0.5pt
   \itemsep 0}
\def\@listii{\leftmargin\leftmarginii
   \labelwidth\leftmarginii\advance\labelwidth-\labelsep
   \topsep 2pt plus 1pt minus 0.5pt
   \parsep 1pt plus 0.5pt minus 0.5pt
   \itemsep 0}
\def\@listiii{\leftmargin\leftmarginiii
    \labelwidth\leftmarginiii\advance\labelwidth-\labelsep
    \topsep 1pt plus 0.5pt minus 0.5pt
    \parsep \z@ \partopsep 0.5pt plus 0pt minus 0.5pt
    \itemsep 0}
\def\@listiv{\leftmargin\leftmarginiv
     \labelwidth\leftmarginiv\advance\labelwidth-\labelsep}
\def\@listv{\leftmargin\leftmarginv
     \labelwidth\leftmarginv\advance\labelwidth-\labelsep}
\def\@listvi{\leftmargin\leftmarginvi
     \labelwidth\leftmarginvi\advance\labelwidth-\labelsep}

\onlineid{0}

\vgtccategory{Research}

\vgtcpapertype{please specify}

\title{When Refreshable Tactile Displays Meet Conversational Agents: Investigating Accessible Data Presentation and Analysis\\ with Touch and Speech}

\author{%
  Samuel Reinders, 
  Matthew Butler, 
  Ingrid Zukerman,
  Bongshin Lee,
  Lizhen Qu, and
  Kim Marriott
}

\authorfooter{
  \item
  	Samuel Reinders, Matthew Butler, Ingrid Zukerman, Lizhen Qu, and Kim Marriott are with Monash University.
  	E-mail: first.last@monash.edu
  \item
  	Bongshin Lee is with Yonsei University,
  	E-mail: b.lee@yonsei.ac.kr
}

\abstract{Despite the recent surge of research efforts to make data visualizations accessible to people who are blind or have low vision (BLV), how to support BLV people's data analysis remains an important and challenging question. As refreshable tactile displays (RTDs) become cheaper and conversational agents continue to improve, their combination provides a promising approach to support BLV people's interactive data exploration and analysis. To understand how BLV people would use and react to a system combining an RTD with a conversational agent, we conducted a Wizard-of-Oz study with 11 BLV participants, where they interacted with line charts, bar charts, and isarithmic maps. Our analysis of participants' interactions led to the identification of nine distinct patterns. We also learned that the choice of modalities depended on the type of task and prior experience with tactile graphics, and that participants strongly preferred the combination of RTD and speech to a single modality. In addition, participants with more tactile experience described how tactile images facilitated a deeper engagement with the data and supported independent interpretation. Our findings will inform the design of interfaces for such interactive mixed-modality systems.}

\keywords{Accessible data visualization, refreshable tactile displays, conversational agents, interactive data exploration, Wizard of Oz study, people who are blind or have low vision.}

\teaser{
  \centering
    \includegraphics[alt={Three images are included in this figure. The first, (a) shows an image of a bar chart that has been authored for display on a RTD. The second, (b) shows a photo of a RTD with that same bar chart displayed on it. The third, (c) shows a users hand touching the bar chart displayed on the RTD, with speech bubbles highlighting an interaction. The user asks the system -- what is the trend of this graph -- while the system responds -- the water storage begins at 83 percent. The three images are used to showcase how a system that combines refreshable tactile displays and conversational agents might be used to assist blind users with data understanding.},width=\linewidth]{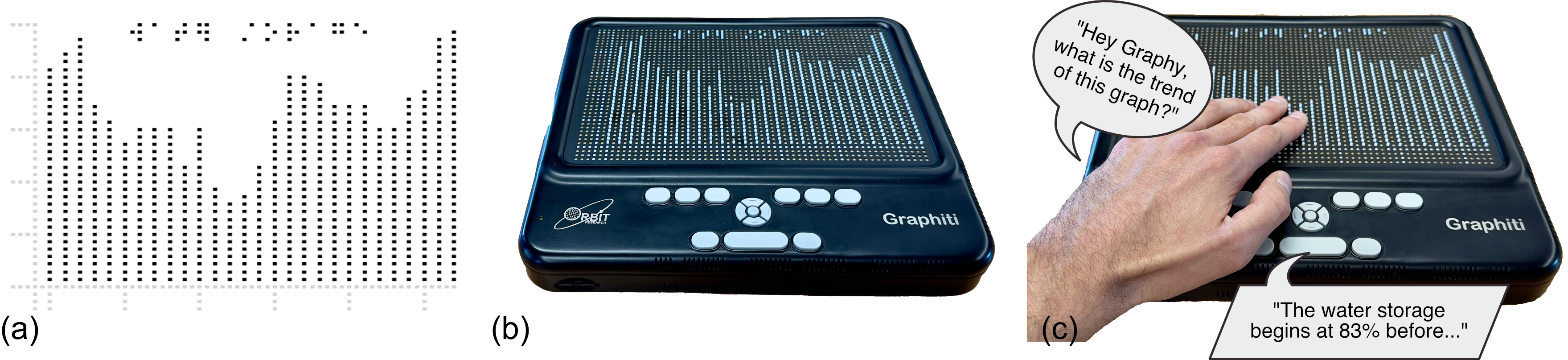} 
    \caption{Using a Wizard-of-Oz method, we explored how refreshable tactile displays (RTDs) can be combined with conversational agents to assist people who are blind or have low vision (BLV) in undertaking data analysis activities. (a)~Bar chart showing water storage over a 27-year period; (b)~That same bar chart rendered on the RTD; and (c)~BLV users could touch the RTD, perform touch gestures, and ask the conversational agent questions to aid their data understanding.}
  \label{fig:teaser}
}

\graphicspath{{figs/}{figures/}{pictures/}{images/}{./}} 

\begin{document}


\firstsection{Introduction}

\maketitle

\label{section:intro}
A recent focus of data visualization research has been on how to best support people who are blind or have low vision (BLV) in their efforts to access visualizations. The research community has paid special attention to automated descriptions of visualizations and conversational interfaces~\cite{SharifEtAlCHI2022,alam2023seechart,Kim2023}, while the sonification community has focused on using non-speech audio to provide accessible data representations for BLV people~\cite{siu2022supporting,thompson2023chart}.

Raised line drawings, called tactile graphics, are widely used in the education of BLV students and in orientation and mobility (O\&M) training. However, traditional printed tactile graphics are expensive to produce and unsuited for interactive usage such as data visualization and analysis. Fortunately, this is about to change, as low-cost refreshable tactile displays (RTDs) come onto the market~\cite{spotlighting2024}. RTDs are the equivalent of a computer display for BLV people. RTDs eliminate the recurring costs associated with traditional tactile graphics: the presentation of a tactile graphic is essentially free after the initial purchase of the device. Furthermore, RTDs support the interactive display of tactile graphics, as an image can be refreshed in only a few seconds. 

However, the use of an RTD in isolation has disadvantages. RTDs currently have low resolution (making it difficult to provide braille labels), tactile graphics without an accompanying textual or verbal description are difficult to comprehend~\cite{BANA2010guidelines}, and there are large disparities in the level of tactile literacy in the BLV community. This suggests that a conversational interface may complement an RTD, and that the combination of an RTD and conversational agent might overcome many of the current barriers BLV people face when analyzing data. This combination, however, has not been previously considered.

Our principal contribution is an initial understanding of how BLV people would use and react to such a multimodal system, from a Wizard-of-Oz study conducted with 11 BLV participants. In our study, participants were introduced to an RTD and were told that they could use direct touch input and gestures to control the RTD and engage in speech interactions with a virtual assistant (i.e., a text-to-speech interface managed by the wizard). They were given three data analysis scenarios, one for training, and completed several tasks in each scenario, which they could address using the RTD and/or the speech interface as they wished. As these scenarios focused on basic data understanding and analysis, we chose to present time series data using line charts and bar charts, and spatial data using isarithmic maps. From the analysis of participant behaviors, we identified preferred modalities and patterns of interaction and how these were influenced by participants' tactile experience and task type (\eg\ identify trend, find values).

The main findings of our study are as follows:
\begin{itemize}
    \item Almost all of our participants indicated that using a combination of an RTD and conversational agent to analyze data had benefits over single formats (\ie\ RTD only or agent only) or traditional tactile graphics alone (Section~\ref{subsection:multimodalaccess}).
    \item Depending on the task type, participants gravitated towards different patterns of interactions. For example, touch was crucial during initial exploration, while touch gestures and speech were more frequently used when identifying values or extrema (Section~\ref{subsection:interactionpatterns}).
    \item Participants' level of tactile experience influenced their interaction patterns (Section~\ref{subsection:tactileexperience}), and more experienced tactile graphic users spoke of how the RTD enabled a deeper engagement with the data and independent interpretation (Section~\ref{subsection:Interview}).
\end{itemize}

Our findings highlight the desire by BLV people to engage in independent and meaningful data exploration, and indicate that this is supported by the combination of an RTD and conversational agent. Our work provides the basis for the future development of such systems.

\section{Related Work}
\label{section:related}
\subsection{Accessible Data Visualization}
The growing availability of data has allowed decision-making to become increasingly data-driven in virtually all professions. Employees are now expected to use tools like Excel or Tableau for data-driven decision making, and to employ a mix of analytics and visualization. The use of data visualization is, however, problematic for BLV people, and we are witnessing a surge of research interest in investigating accessible alternatives (\eg\ \cite{lee2020reaching,kim2021accessible,marriott2021inclusive,dagstuhl2023inclusive,Fan2022,seo2024,zhang2024}). 

One approach is to provide a written or spoken description of a graphic. Researchers have investigated what makes a good description~\cite{jung2021communicating}, how to automatically generate a description from a graphic or answer questions about a graphic (see~\cite{farahani2023automatic,HoqueEtAl2022chart} for recent reviews), or how to generate descriptions directly from data~\cite{YuReiterEtAlNLE2007,PuduppullyEtAlACL2019,RebuffelEtAl2020}. Some researchers have explored how to incorporate these capabilities into systems designed to support data analysis by BLV people~\cite{SharifEtAlCHI2022,alam2023seechart,Kim2023}.

Another approach is to use non-speech audio, known as sonification, to convey information. For instance, the SAS Graphics Accelerator uses sonification and tables to provide access to visualizations~\cite{SASGraphicsAccelerator}. Sonification can be used as an alternative format to data visualizations, or alongside data visualizations~\cite{Kramer2010}. Recently, there has been interest in combining speech with sonification~\cite{holloway2022infosonics,siu2022supporting,thompson2023chart}.

The third approach is to provide raised line drawings known as tactile graphics, which are encountered by many BLV students at school. Transcription guidelines recommend their use for accessible provision of maps, diagrams, and graphs, rather than textual or verbal descriptions, in particular if the reader needs the information to participate in discussions or complete a task~\cite{BANA2010guidelines}. Most tactile graphics are produced using swell paper or embossing~\cite{rowell2003world}. However, printed tactile graphics are not well-suited for interactive data exploration, because of the cost and speed of production. Therefore, there has been considerable research into displays that combine auditory and tactile feedback as alternatives to tactile graphics, which include touch screens with audio labels and with/without vibration feedback (\eg~\cite{goncu2011gravvitas,Miele2006}). However, previous research suggests that exploration is faster and more accurate with tactile graphics than with touchscreen-based approaches~\cite{butler2021technology}.

\subsection{Refreshable Tactile Displays}
\label{section:RTD}
A promising alternative is large refreshable tactile displays (RTDs)~\cite{vidal2007graphical,Yang2021}. Most consist of a grid of pins controlled by electro-mechanical actuators~\cite{Yang2021}. Their great advantage over traditional tactile graphics is that it takes only a few seconds to display a new graphic. In other words, an RTD is the tactile equivalent of a computer monitor, and like a monitor, it supports interactive data exploration. 

Until recently, RTDs have been prohibitively expensive. However, the DAISY Consortium's Transforming Braille Project has spurred the development of cheaper devices, which are now entering the market. These include the DotPad~\cite{DotInc} and Monarch~\cite{APH}, which provide 2,400 (60$\times$40) and 3,840 (96$\times$40) pins. Other devices on the market include the Graphiti~\cite{Orbit} (2,400 (60$\times$40) pins) and Metec~\cite{Metec} (6,240 ($104\times 60$) pins), which offer several pin heights and touch-sensitive pins.

Research into the use of RTDs has focused on the display of static images in the fields of art~\cite{Gyoshev2018} and book diagrams and illustrations~\cite{Kim2019,Namdev2015,Park2016}. Other work has looked at the display of dynamic images, such as sports matches~\cite{Ohshima2021football} and animations~\cite{Holloway2022}. RTDs have also been used to show maps for building a cognitive model pre-travel~\cite{brayda2019refreshable,Ivanchev2014,Motoyoshi2018,Schmitz2012,Zeng2015}, or as mobility aids updated in real-time to show position/obstacles as a BLV person moves through an environment~\cite{Brayda2018updated,Zeng2012,zeng2014examples}. 

There has been almost no research into the use of RTDs for interactive data visualization. One paper describes the co-design of tools for the interactive exploration of set diagrams and parallel vectors on an RTD~\cite{elavsky2023data}, and a poster at VIS 2023 examined stakeholder perspectives of how RTDs could be used to improve access to graphics, including data graphics~\cite{holloway2024refreshable}. There has been more research into systems that use RTDs with audio labels to display maps automatically generated from web-based map systems, which support search, zooming, panning, and route finding (\eg~\cite{Schmitz2012,zeng2014examples,Zeng2015}). 

\subsection{Multimodal Display of Graphical Information}
\label{section:multimodal}
A limitation of RTDs is their low resolution when compared with traditional tactile graphics~\cite{OModhrain2015,Holloway2022}. This makes the provision of braille labels problematic, and is one of the reasons why audio labels are provided on most RTD map applications (\eg~\cite{Schmitz2012,zeng2014examples,Zeng2015}). Indeed, in their review, Butler \etal~\cite{butler2021technology} found that audio labels on tactile graphics are generally preferred to braille labels and lead to faster performance. Furthermore, it is recommended that tactile graphics be accompanied by a spoken or written description~\cite{BANA2010guidelines}.

Advances in question-answering interfaces and conversational agents suggest that we can go one step further and combine an RTD with a conversational agent. We know of no prior research into such a combination. However, there has been some research into combining conversational agents with tactile maps~\cite{cavazos2019jido}, tactile graphics~\cite{Fusco2015}, art pieces~\cite{Bartolome2019,Quero2018}, and with 3D printed models~\cite{Shi2019,Reinders2020,Reinders2023}. Shi \etal~\cite{Shi2019} determined that in addition to audio labels, 3D printed models should allow BLV users to ask questions using conversational language. Reinders \etal~\cite{Reinders2020} found that BLV participants desire multimodal interactions that combine spoken dialogue, touch gestures, and haptic vibratory feedback; this finding was followed by design recommendations that heavily focused on conversational interactions~\cite{Reinders2023}.

\subsection{Wizard-of-Oz Method}
\label{subsection:WoZ}
A Wizard-of-Oz (WOz) study~\cite{dahlback1993wizard} is a research method where an end user interacts with an interface that, to some degree, is being operated by a `wizard' who fulfills an interaction functionality that is yet to be fully implemented~\cite{kelley1984}. WOz studies often involve eliciting interactions from end users, with the wizard simulating system behaviors that complete their interactions~\cite{Wobbrock2009,Connell2013,Shi2017b}. WOz enables researchers to observe the use of speculative interfaces or artefacts, the outcomes of which can influence future designs.

WOz studies have been used in various domains and contexts. Salber and Coutaz~\cite{salber2005} demonstrated how WOz can be extended to the analysis of multimodal interfaces, and formulated a set of requirements for a generic multimodal WOz platform. Kahn \etal~\cite{kahn2008} employed a WOz technique to control some of a humanoid robot's speech and actions to investigate patterns in human-robot interactions. In information visualization research, Walny \etal~\cite{walny2012understanding} exploited the WOz method to investigate pen and touch interactions for chart creation and manipulation on interactive whiteboards.
Recently, researchers applied WOz to investigate BLV users preferred interaction preferences with multimodal systems combining 3D printed models and conversational agents~\cite{Shi2017b,Reinders2020}. Kim \etal~\cite{Kim2023} conducted a WOz study to understand if and how Question Answering systems can help BLV users' visualization comprehension.  

Our WOz study explores a system that combines a conversational agent with an RTD to support BLV people's interactive data analysis. Our objective is to propose initial design guidelines and to meaningfully engage BLV people in the design of such a system.

\begin{figure*}[t]
\begin{center}
\begin{tabular}{ccccc}
    \includegraphics[alt={(a) shows a line graph depicting the price of a stock increasing over time. It has axes and a braille title.},width=5cm]{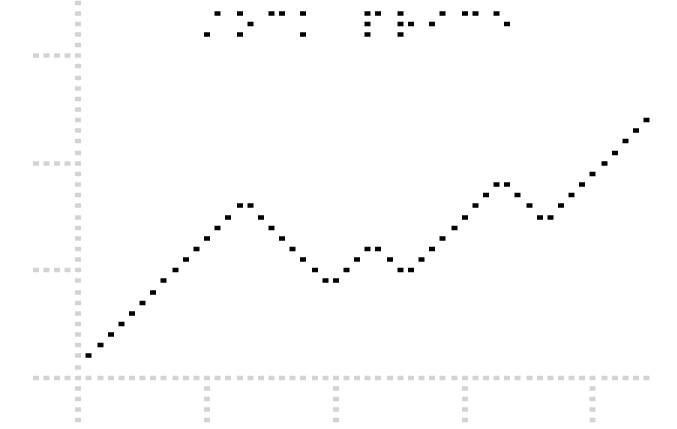} & ~~~~ &
    \includegraphics[alt={(b) is a line graph showing water storage data. It has axes, a braille title and a legend label -- water.},width=5cm]{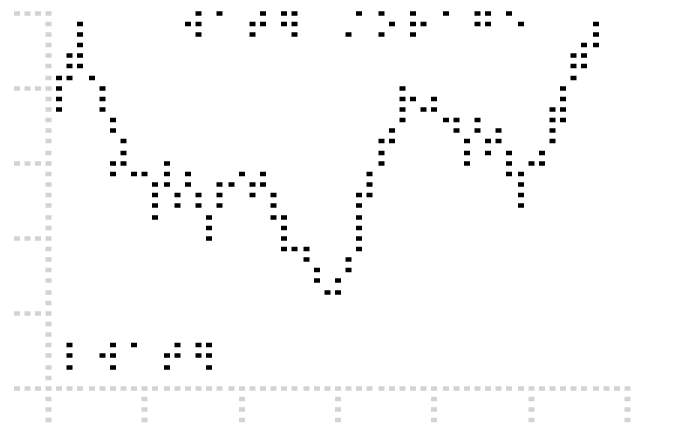} & ~~~~ &
    \includegraphics[alt={(c) is a line graph showing water rainfall data. It has axes, a braille title and a legend label -- rain.},width=5cm]{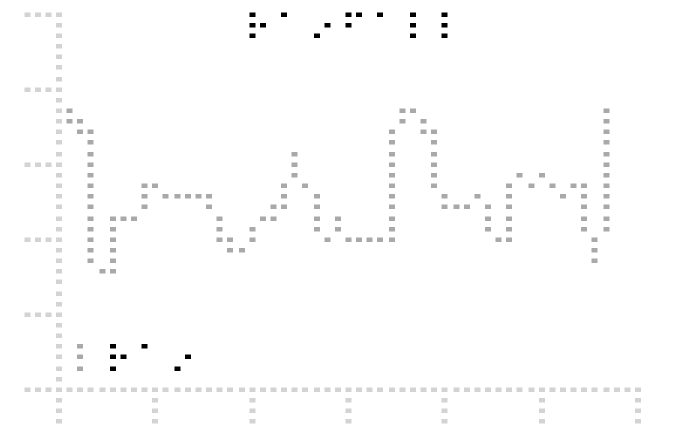}\vspace*{2mm}\\
    (a)~{\Stock\ line graph} & & (b)~{\Water\ storage line graph} & & (c)~{\Water\ rainfall line graph}\vspace*{-3mm} \\
    \vspace*{5mm} \\
    \includegraphics[alt={(d) is a multi line graph that shows water storage and water rainfall overlapping. It has axes, a braille title and two legend labels -- rain and water.},width=5cm]{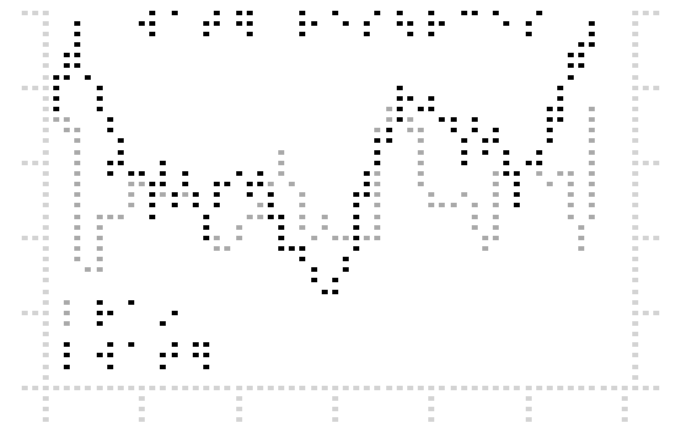} & ~~~~ &
    \includegraphics[alt={(e) showcases water storage data transformed into a bar graph. It has axes and a braille title.},width=5cm]{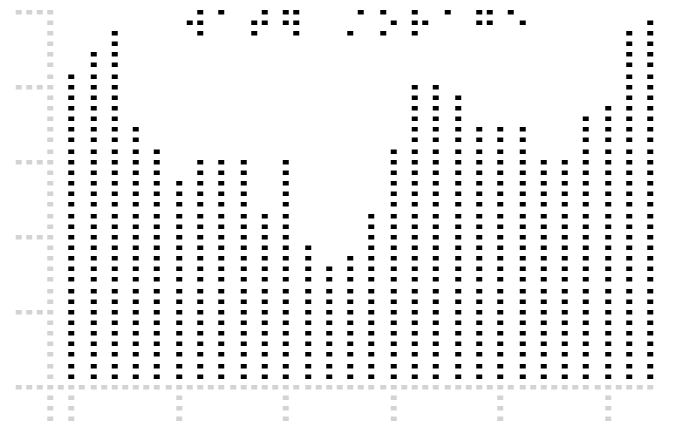} & ~~~~ &
    \includegraphics[alt={(f) is an isarithmic map of Australia graphing fire weather data. It has a braille title and four legend labels.},width=5cm]{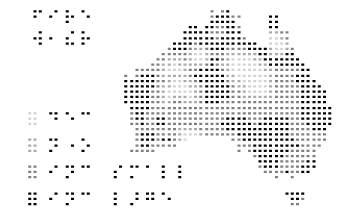}\vspace*{2mm}\\
    (d)~{\Water\ storage and rainfall multi line graph} & & (e)~{\Water\ storage bar graph} & & (f)~{\Fire\ isarithmic map}\vspace*{-3mm} \\
\end{tabular}
\end{center}
\caption{Six of the data visualizations used in the user study. For \Water, participants could choose between rainfall or water storage data, or both together. The individual \Water\ line graphs could also be presented using bar graphs (\eg\ (e)) that represent the same information. \label{tab:dataCharts}}
\end{figure*}
\section{User Study -- Methodology}
\label{section:method}
We conducted a user study (1)~to evaluate the perceived usefulness of the combination of an RTD and a conversational agent to support data analysis by BLV people and (2)~to identify the ways in which BLV people wish to interact with such a combination for data analysis. A particular focus was on understanding how they wished to use the different modalities of touch, touch gesture, and speech.

To achieve these goals, we used a WOz method to simulate system behaviors and elicited user interaction needs. This allowed us to capture how the BLV participants chose to interact with an RTD and conversational agent when conducting data analysis free of technological constraints. We also conducted a questionnaire and semi-structured interview to further explore participant responses to the system.

\subsection{Participants}
\label{subsection:participants}
We recruited 11 BLV participants (Table~\ref{tab:participants}) from our lab-managed participant contact pool and through mailing lists of BLV support and advocate groups. Participants were aged between 21 and 77 (\textit{$\mu$} = 44, \textit{$\sigma$} = 17.2). Eight participants identified as totally blind, while three were legally blind. Onset of vision loss was congenital ($n=7$) or occurred during childhood ($n=1$), adolescence ($n=2$), or adulthood ($n=1$). Participants varied in their experience with tactile graphics: five reported substantial exposure to and confidence using tactile graphics, and three reported some use but lacked confidence, while others reported either limited ($n=1$) or no exposure ($n=2$). 

\begin{table}[t]
\caption{Participant demographic information, detailing age, level of blindness (TB = totally blind, LB = legally blind), onset of vision loss, and self-rated competency using Tactile Graphics (TG).\label{tab:participants}}
\small
\centering
\begin{tabular}{|c|c|c|c|c|}\hline
\textbf{Part.} & \textbf{Age} & \textbf{Blindness} & \textbf{Onset} & \textbf{TG Experience}\\ \hline
P1  & 21 & LB   & Congenital & Substantial\\
P2  & 25 & TB   & Congenital & Substantial\\
P3  & 42 & TB   & 5 y.o.        & Substantial\\
P4  & 54 & TB   & Congenital & None\\
P5  & 33 & LB   & 15 y.o.        & Limited\\
P6  & 53 & TB   & Congenital & Some\\
P7  & 45 & LB   & 15 y.o.        & Some\\
P8  & 40 & TB   & Congenital & Some\\
P9  & 77 & TB   & 45 y.o.        & None\\
P10 & 65 & TB   & Congenital & Substantial\\
P11 & 29 & TB   & Congenital & Substantial\\ \hline
\end{tabular}
\vspace*{-1mm}
\end{table}

All participants regularly used conversational interfaces, including Google Assistant ($n=10$), Siri ($n=9$), and Alexa ($n=5$). These interfaces were commonly accessed on smartphones ($n=11$) and smart speakers ($n=10$), followed by smart watches ($n=7$), computers ($n=7$), tablets ($n=5$), and television ($n=3$). Additionally, several participants ($n=5$) mentioned some use of ChatGPT.

Participants self-reported their technology comfort level using the Technology Adoption Cycle scale~\cite{Rogers2003}. Most ($n=10$) were Early Majority and one was Late Majority. All other categories, Innovator, Early Adopter, and Laggard, were absent.

Participants were also asked to self-report their level of comfort with basic and complex data manipulation/analysis tasks on a five-point Likert scale. Most participants were comfortable with basic tasks (\eg\ creating a budget):  Strongly Agree ($n=5$), Agree ($n=4$), and Neutral ($n=2$). Comfort with complex tasks (\eg\ performing a statistical analysis) was more mixed: Strongly Agree ($n=1$), Agree ($n=4$), Neutral ($n=3$), Disagree ($n=1$), and Strongly Disagree ($n=2$).

\subsection{Materials}
For the study we used Graphiti~\cite{Orbit} (Figure~\ref{fig:teaser}(b)). The Graphiti's display has 2,400 (60$\times$40) pins, each of which can be actuated to one of four heights. The Graphiti includes keys to navigate between graphics, and the display can render Braille text. When connected to a computer, a PC application can be used to create or adjust graphics, with changes appearing in real-time on the RTD.

As our work is the first utilizing an RTD for data presentation, we decided to focus on basic data understanding. We selected univariate time series and spatial data, and visualization types we believed our participants would be most familiar with. We selected datasets from finance and climate--\Stock, \Water\, and \Fire--that we believed would be both timely and of interest to participants~\cite{Kim2023}, and created several visualizations (line charts, bar charts, and an isarithmic map)~(Figure~\ref{tab:dataCharts}). 

\begin{itemize}
    \myitem \Stock\ gives the price of a stock over four and a half years (Figure~\ref{tab:dataCharts}(a)). It was presented using a line chart, with years on the horizontal axis and price on the vertical axis. This was shown first to familiarize participants with the device and its capabilities. 
    
    \myitem \Water\ gives the rainfall and water storage levels in the state of \State, \Country\ between the years 1995 and 2022. The water storage and rainfall information were presented as line charts, either separately (Figure~\ref{tab:dataCharts}(b)-(c)) or together as a multi-line chart with dual axes (Figure~\ref{tab:dataCharts}(d)), depending on the participants' preference.\footnote{We are aware that dual axes time series may be visually misleading, though for a counter-argument see~\cite{DualAxes}. However, the limited resolution of the RTD effectively precludes the use of line chartss stacked above one another, which is the most commonly recommended alternative. We therefore decided to provide our participants with the choice.}  
    
    Participants could also request to change the presentation format from line chart to bar chart (Figure~\ref{tab:dataCharts}(e)). Line charts are better suited to showing trends, while bar charts are better suited to discrete comparisons. We therefore decided to allow participants to change the visual encoding. 
    
    \myitem \Fire\ gives changes in the number of days with dangerous fire weather across Australia between the years 1951 and 2022, using an isarithmic map (Figure~\ref{tab:dataCharts}(f)). Data points are charted using four pin heights on a map of the country: Height~1 (the shallowest) represents a yearly decrease in fire weather, Height~2 represents no change, Height~3 a yearly increase of between 1 and 10 days, and Height~4 (the tallest) an increase of between 11 and 20 days.
\end{itemize}

All materials were developed with feedback from an experienced tactile transcriber. All visualizations included a braille title and legend. Due to the limited resolution of the Graphiti (in common with other RTDs), braille axis labels and markers were not provided.

\subsection{Study Setup and Procedure}
User study sessions took place at a location convenient to each participant, either at their residence or at the researchers' university campus. One or two researchers were present during the experiment. 

\subsubsection{The Role of the Wizard}
A member of the research team took the role of the wizard during user study sessions. The wizard acted on behalf of the RTD device and conversational agent in response to participants' interactions. The wizard used the Graphiti PC application to adjust graphics in real-time (\eg\ zooming or changing which visualization was displayed). To provide feedback to participants quickly and consistently, the wizard answered questions posed to the conversational agent by selecting responses from a script created for each visualization. In line with Kim \etal's work~\cite{Kim2023}, each script covered the visualization title, overview, axes values, data values for each data point, trend, extrema (minimum and maximum), and background information regarding the visualizations (\eg\ ``what is a bar chart,'' ``what constitutes a dangerous fire weather day''). Responses were generated on the fly using text-to-speech, and played back using a connected speaker. If a particular request was outside of the scope of a script and could not be satisfied (\eg\ converting a line chart into a scatter plot), the wizard would inform the participant that the functionality in question had not been implemented and would respond: \textit{``I currently don't understand how to answer that, as I am a prototype, but I will note your request down for the future.''}

In line with typical WOz studies, we informed our participants of the true nature of the WOz deception at the conclusion of the study session. However, two participants (P2 and P3, having had previous experience with WOz studies) asked during the study if the interface was ``real,'' and they were informed of the true nature of the interface before continuing with the study.

\subsubsection{Survey and Overview}
Sessions began with a brief questionnaire that included questions about participant demographic information, accessible format use, technology use, and data proficiency.  Then, participants were informed that they would be interacting with a prototype device that combines an RTD and a conversational agent to help them explore data and complete basic data analysis tasks. Participants were asked to verbalize any actions they wanted to perform using the `think-aloud' protocol, as the focus of the session was on observing their chosen ways of interacting with the device. This allowed the wizard to act on behalf of the RTD and conversational agent when necessary. Think-aloud has been used widely in elicitation studies~\cite{Wobbrock2009}, including those that utilize WOz experiments in the accessibility space~\cite{Shi2017b,Reinders2020,Kim2023}.

\subsubsection{Familiarization with the System}
The \Stock\ visualization was used to familiarize participants with the Graphiti device, the think-aloud protocol, and the ways they might interact with the device. The researcher first provided an overview of the device and its technical capabilities. This included outlining that it could generate graphics by raising small pins arranged in 40 rows of 60 pins, and that it supported natural language understanding and generation. Participants were provided with example interactions they might choose to use, \eg\ touch gestures, asking questions with speech using the wake phrase \textit{``Hey Graphy,''} and using the RTD's buttons, either in isolation or in combination. Participants were then asked to explore the \Stock\ visualization.

Once they finished their exploration of the \Stock\ visualization, participants were directly elicited to choose modalities and interaction techniques  to perform a series of basic tasks: (1)~getting the title of a graphic, (2)~getting an in-depth description of a graphic, and (3)~selecting a data point and extracting its value (detail on demand).  The wizard acted on behalf of the RTD to fulfill the participants' requests.

\subsubsection{Interacting with \Water\ and \Fire\ Visualizations}
Participants were first introduced to the \Water\ visualization, followed by the \Fire\ visualization.

\bipstart{Water} The researcher explained that participants could choose to display water catchment storage, rainfall, or both types of data together: the appropriate single or multi-line chart was displayed on the RTD based on the participants' choice.

Participants were then asked to explore \Water. Upon the first touch, the agent provided a spoken overview of the \Water\ graph. Participants were also invited to interact with the visualization to perform the following operations: (1)~zooming, (2)~panning, (3)~adding/removing data sets, (4)~filtering, (5)~transforming to another representation (line chart to bar chart), and (6)~undo. Again, elicited interactions were fulfilled by the wizard.

Participants were then asked to complete several data analysis tasks using either the storage line chart, rainfall line chart, their respective bar charts, or the combined storage and rainfall multi-line chart. 

\bipstart{Fire} Participants were asked to explore the \Fire\ isarithmic map visualization, with the agent providing a spoken overview when first touched. Participants were then asked to perform similar analysis tasks to those completed with \Water. 

\bipstart{Analysis tasks} For both \Water\ and \Fire, participants were asked to complete three types of tasks, representative of identification, comparison, and summarization~\cite{Munzner2015}. Our choice of tasks was also influenced by other systems aimed at facilitating data understanding by BLV users~\cite{SharifEtAlCHI2022,HoqueEtAl2022chart}.

\begin{itemize}
    \myitem \textit{Identifying specific values}: for the \Water\ visualization, this task involved reporting the most recent storage/rainfall, and that in 2005, and for \Fire, this task required identifying the number of fire-weather days in \City, \Country.
    
    \myitem \textit{Identifying extreme values}: for the \Water\ visualization, this task involved reporting on the lowest and highest storage/rainfall, and for \Fire, participants had to find an area where the number of fire-weather days had decreased.

    \myitem \textit{Describing the trend}: this task required participants to report on the general trend of the phenomenon of interest in both visualizations.
\end{itemize}
     
Participants were asked to complete these tasks using any interactions or transformation operations they found suitable. Afterwards, they were given the opportunity to further explore \Water\ and \Fire, and asked to articulate any interesting facts that they had identified.

\subsubsection{Interview}
The session ended with a semi-structured interview. The researchers asked questions to better understand why participants used particular interaction modalities, their comfort level during interactions, particular use cases, whether the combination of an RTD and conversational agent would make them more confident undertaking data tasks, and if the combination was preferred to a conversational agent, RTD, or tactile graphic in isolation.  Sessions lasted between 1.5 to 2 hours, and participants were compensated with a \$100 AUD gift card.

\subsection{Data Collection and Analysis}
\label{subsection:data}
User study sessions were video and audio recorded. All captured footage was transcribed.

Our aim was to capture and categorize the way in which our BLV participants chose to interact with the system. To achieve this goal, we classified each interaction based on the combination of modalities used (touch, conversation, or touch gestures), as well as their order and intent. One researcher derived an initial set of patterns by classifying the interactions of the first user study session (P1), which were discussed and confirmed with a second researcher. This set was supplemented with additional patterns identified in subsequent user study sessions (P2 - P11). Whenever new patterns were added, all previous sessions were reassessed. On completion, the second researcher validated a subset of the classified interactions, with a subsequent meeting held to reconcile conflicts. The final set of interaction patterns is presented in Section~\ref{section:results}.

A thematic analysis was performed similarly on (1)~comments made by participants during a session and (2)~participant responses to the semi-structured interview. An initial set of codes was devised based on the researchers' observation of the first user study session (P1). Two members of the research team independently coded this session and met to compare codes, which were then refined and extended. All subsequent user study sessions were then coded by one researcher, with the second researcher validating a subset of the coded data. An additional meeting was held to discuss and reconcile any coding conflicts. Whenever new codes were added, combined or modified, all previous sessions were reassessed. The complete set of coded data was examined and analyzed to generate a set of themes. Three researchers met to review the themes, which were then refined and consolidated. The final set of themes is described in Section~\ref{subsection:Interview}.

\section{Results}
\label{section:results}
We present participant command interaction preferences, modality interaction patterns, participant conversational queries, and a thematic analysis of participant responses.

\subsection{Command Interaction Preferences}
\label{subsection:interactionTechniques}
Multiple participants controlled the system (referred to from now on as `Graphy') using commands based on prior experience with devices, including smartphones, smart speakers, and computers. 

To access the visualization title, six participants preferred to touch read a braille label, \eg\ \textit{``I would touch the top and feel for braille''} (P6). Graphy provided an overview of each visualization upon first touch; to repeat the overview, seven participants asked the device, \eg\ \textit{``Hey Graphy, can you describe the chart?''} (P3). 

When extracting detail on demand, participants either performed touch gestures ($n$ = $5$), \eg\ single and double tapping to extract descriptions of values at the location at which the gesture was performed, or combined touch gestures with conversation ($n = 6$), \eg\ tap and hold on a point and ask a question, expecting the device to use their location in conjunction with the question.

\begin{table}[t]
\renewcommand{\arraystretch}{1.3}
\caption{Modalities that participants used when performing basic functions and data operations (T = Touch, C = Conversation, G = Touch Gesture, K = Keyboard, G+C = Touch Gesture and Conversation); modalities used in more than 20\% of each task/operation are highlighted.
\label{tab:interactionTechniques}}
\small
\centering
\begin{tabular}{|lccccc|}
\hline
\multicolumn{1}{|c|}{\multirow{2}{*}{\textbf{Operation}}} & \multicolumn{5}{c|}{\textbf{Modalities}} \\ \cline{2-6} 
\multicolumn{1}{|c|}{} & 
\multicolumn{1}{>{\centering\arraybackslash}p{0.075\columnwidth}|}{T} &  
\multicolumn{1}{>{\centering\arraybackslash}p{0.075\columnwidth}|}{C} & 
\multicolumn{1}{>{\centering\arraybackslash}p{0.075\columnwidth}|}{G} & 
\multicolumn{1}{>{\centering\arraybackslash}p{0.075\columnwidth}|}{K} & 
\multicolumn{1}{>{\centering\arraybackslash}p{0.075\columnwidth}|}{G+C} \\ \hline
\multicolumn{1}{|l|}{Visualization Title} & \multicolumn{1}{c|}{\cellcolor[HTML]{FFCE93}6} & \multicolumn{1}{c|}{\cellcolor[HTML]{FFCE93}3} & \multicolumn{1}{c|}{2} & \multicolumn{1}{c|}{0} & 0 \\ \hline
\multicolumn{1}{|l|}{Visualization Overview} & \multicolumn{1}{c|}{\cellcolor[HTML]{FFCE93}3} & \multicolumn{1}{c|}{\cellcolor[HTML]{FFCE93}7} & \multicolumn{1}{c|}{1} & \multicolumn{1}{c|}{0} & 0 \\ \hline
\multicolumn{1}{|l|}{Detail on Demand} & \multicolumn{1}{c|}{0} & \multicolumn{1}{c|}{0} & \multicolumn{1}{c|}{\cellcolor[HTML]{FFCE93}5} & \multicolumn{1}{c|}{0} & \cellcolor[HTML]{FFCE93}6 \\ \hline
\multicolumn{1}{|l|}{Zooming} & \multicolumn{1}{c|}{0} & \multicolumn{1}{c|}{0} & \multicolumn{1}{c|}{\cellcolor[HTML]{FFCE93}6} & \multicolumn{1}{c|}{1} & \cellcolor[HTML]{FFCE93}4 \\ \hline
\multicolumn{1}{|l|}{Panning} & \multicolumn{1}{c|}{0} & \multicolumn{1}{c|}{\cellcolor[HTML]{FFCE93}4} & \multicolumn{1}{c|}{\cellcolor[HTML]{FFCE93}6} & \multicolumn{1}{c|}{0} & 1 \\ \hline
\multicolumn{1}{|l|}{Adding/Removing Data} & \multicolumn{1}{c|}{0} & \multicolumn{1}{c|}{\cellcolor[HTML]{FFCE93}8} & \multicolumn{1}{c|}{0} & \multicolumn{1}{c|}{\cellcolor[HTML]{FFCE93}3} & 0 \\ \hline
\multicolumn{1}{|l|}{Filtering} & \multicolumn{1}{c|}{0} & \multicolumn{1}{c|}{\cellcolor[HTML]{FFCE93}8} & \multicolumn{1}{c|}{1} & \multicolumn{1}{c|}{2} & 0 \\ \hline
\multicolumn{1}{|l|}{Transforming} & \multicolumn{1}{c|}{0} & \multicolumn{1}{c|}{\cellcolor[HTML]{FFCE93}9} & \multicolumn{1}{c|}{1} & \multicolumn{1}{c|}{1} & 0 \\ \hline
\multicolumn{1}{|l|}{Undo} & \multicolumn{1}{c|}{0} & \multicolumn{1}{c|}{\cellcolor[HTML]{FFCE93}9} & \multicolumn{1}{c|}{1} & \multicolumn{1}{c|}{1} & 0 \\ \hline
\end{tabular}
\end{table}

For zooming, six participants used the pinch-to-zoom gestures used on their smartphones, \eg\ \textit{``I am so used to the iOS gestures, can I do that?''} (P8), while four combined gestures with voice, \eg\ \textit{``Hey Graphy, zoom on the area I am holding''} (P7). For panning, six participants used swiping gestures from smartphones, \eg\ swiping to the left or right with one or more fingers.

For operations that do not cleanly map to smartphone gestures, participants gravitated towards asking the device, or in some cases, using the RTD's keyboard. Most participants filtered, transformed, added/removed data, and reversed operations through conversation, \eg\ \textit{``Hey Graphy, only show areas of the isarithmic map where fire risk has decreased''} (P3). Several participants chose to use the RTD's braille keyboard, some of whom indicated that it would be better to be able to connect a QWERTY keyboard, \eg\ \textit{``I want to type $\ldots$ like when I interact with ChatGPT''} (P6). 

\subsection{Modality Interaction Patterns}
\label{subsection:interactionpatterns}
\begin{figure}[h]
  \centering
  \includegraphics[alt={An image showing two participants interacting with visualizations rendered on the refreshable tactile display. In the first image, a participant has their hands tracing a line chart that has been rendered on the display. In the second image, a participant is touch reading braille labels that belong to a legend associated with an isarithmic map of Australia that is rendered on the display.},width=1.00\columnwidth]
  {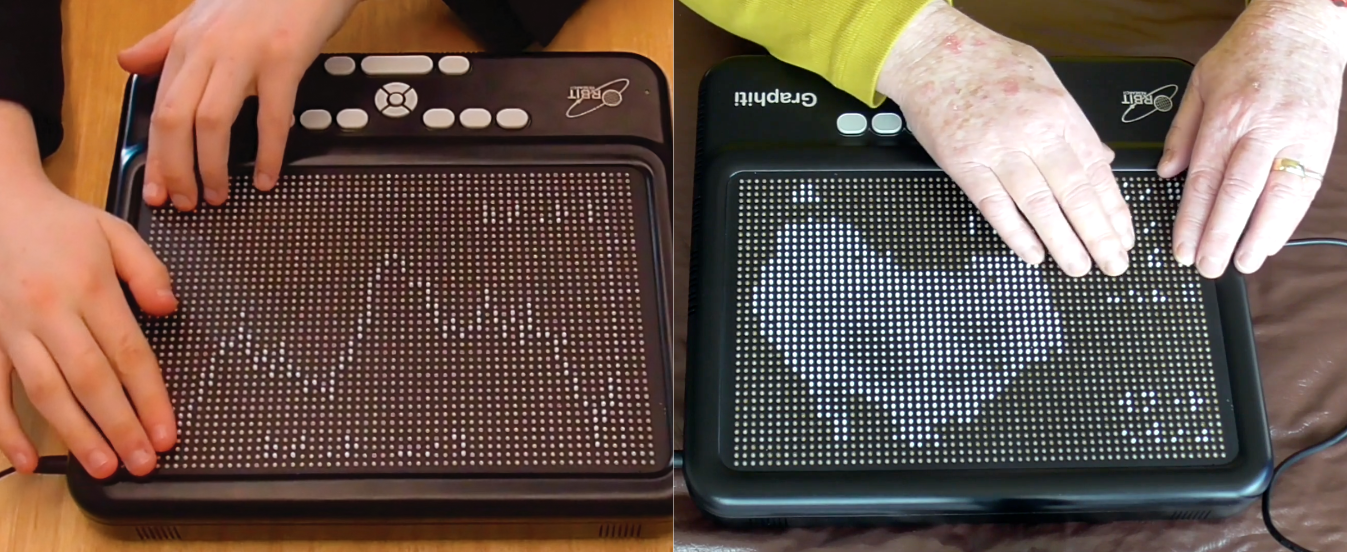}
  \caption{Participants interacting with \Water\ and \Fire\ visualizations.\label{fig:interactions}} 
\end{figure}

All participants combined touch, touch gestures, and natural language queries to interpret and analyze the data presented in the visualizations (Figure \ref{fig:interactions}, Table \ref{tab:interactionTechniques}). 

\begin{table}[t]
\renewcommand{\arraystretch}{1.3}
\caption{Modality interaction patterns specified; the modalities involved in each pattern (T = Touch, G = Touch Gestures, C = Conversation).}\label{tab:interactionPatterns}
\small
\centering
\begin{tabular}{|p{0.35\textwidth}|ccc|}
\hline
\multicolumn{1}{|c|}{\multirow{2}{*}{\textbf{Modality Interaction Pattern}}} &
  \multicolumn{3}{l|}{\textbf{Modalities}} \\ \cline{2-4} 
  \multicolumn{1}{|c|}{} &
  \multicolumn{1}{c|}{T} &
  \multicolumn{1}{c|}{C} &
  G \\ \hline
Conversation Only &
  \multicolumn{1}{c|}{} &
  \multicolumn{1}{c|}{*} &
   \\ \hline
Conversation and Undirected Touch &
  \multicolumn{1}{c|}{*} &
  \multicolumn{1}{c|}{*} &
   \\ \hline
Conversation followed By Directed Touch &
  \multicolumn{1}{c|}{*} &
  \multicolumn{1}{c|}{*} &
   \\ \hline
Directed Touch Only &
  \multicolumn{1}{c|}{*} &
  \multicolumn{1}{c|}{} &
   \\ \hline
Directed Touch followed By Conversation &
  \multicolumn{1}{c|}{*} &
  \multicolumn{1}{c|}{*} &
   \\ \hline
Directed Touch followed By Touch Gesture &
  \multicolumn{1}{c|}{*} &
  \multicolumn{1}{c|}{} &
  * \\ \hline
Interleaved Directed Touch and Conversation &
  \multicolumn{1}{c|}{*} &
  \multicolumn{1}{c|}{*} &
   \\ \hline
Interleaved Directed Touch and Touch Gesture &
  \multicolumn{1}{c|}{*} &
  \multicolumn{1}{c|}{} &
  * \\ \hline
Interleaved Directed Touch, Conversation, and Touch Gesture &
  \multicolumn{1}{c|}{*} &
  \multicolumn{1}{c|}{*} &
  * \\ \hline
\end{tabular}
\end{table}

\setlength{\tabcolsep}{3.7pt}
\begin{table*}[t]
\caption{Modality interaction patterns for four task types (exploration, identifying specific values, identifying extreme values, and describing a trend); Touch is directed unless otherwise specified; number of times each pattern was used, percentage for each task type, and number of unique participants that performed each pattern during each task type; finally, the top two patterns employed during each task type are highlighted.\label{tab:interactionPatternsUse}}
\small
\centering
\begin{tabular}{|l|cccccccccccc|}
\hline
\multicolumn{1}{|c|}{} &
  \multicolumn{12}{c|}{\textbf{Task Types}} \\ \cline{2-13} 
\multicolumn{1}{|c|}{\multirow{2}{*}{\textbf{Interaction Modality Pattern}}} &
  \multicolumn{3}{c|}{\textbf{Exploration}} &
  \multicolumn{3}{c|}{\textbf{Specific Values}} &
  \multicolumn{3}{c|}{\textbf{Extreme Values}} &
  \multicolumn{3}{c|}{\textbf{Trend}} \\
\multicolumn{1}{|c|}{} &
  \# times &
  \% &
  \multicolumn{1}{c|}{\# participants} &
  \# times &
  \% &
  \multicolumn{1}{c|}{\# participants} &
  \# times &
  \% &
  \multicolumn{1}{c|}{\# participants} &
  \# times &
  \% &
  \# participants \\ \hline
Conversation Only &
  0 &
  0\% &
  \multicolumn{1}{c|}{0} &
  2 &
  6\% &
  \multicolumn{1}{c|}{2} &
  3 &
  9\% &
  \multicolumn{1}{c|}{2} &
  2 &
  10\% &
  1 \\
Conversation and Undirected Touch &
  0 &
  0\% &
  \multicolumn{1}{c|}{0} &
  2 &
  6\% &
  \multicolumn{1}{c|}{2} &
  3 &
  9\% &
  \multicolumn{1}{c|}{3} &
  1 &
  5\% &
  1 \\
Conversation followed By Touch &
  0 &
  0\% &
  \multicolumn{1}{c|}{0} &
  1 &
  3\% &
  \multicolumn{1}{c|}{1} &
  0 &
  0\% &
  \multicolumn{1}{c|}{0} &
  \cellcolor[HTML]{FFCE93}4 &
  \cellcolor[HTML]{FFCE93}19\% &
  \cellcolor[HTML]{FFCE93}3 \\
Touch Only &
  \cellcolor[HTML]{FFCE93}16 &
  \cellcolor[HTML]{FFCE93}49\% &
  \multicolumn{1}{c|}{\cellcolor[HTML]{FFCE93}8} &
  0 &
  0\% &
  \multicolumn{1}{c|}{0} &
  2 &
  6\% &
  \multicolumn{1}{c|}{2} &
  \cellcolor[HTML]{FFCE93}9 &
  \cellcolor[HTML]{FFCE93}43\% &
  \cellcolor[HTML]{FFCE93}6 \\
Touch followed By Conversation &
  0 &
  0\% &
  \multicolumn{1}{c|}{0} &
  \cellcolor[HTML]{FFCE93}8 &
  \cellcolor[HTML]{FFCE93}24\% &
  \multicolumn{1}{c|}{\cellcolor[HTML]{FFCE93}6} &
  \cellcolor[HTML]{FFCE93}7 &
  \cellcolor[HTML]{FFCE93}21\% &
  \multicolumn{1}{c|}{\cellcolor[HTML]{FFCE93}6} &
  1 &
  5\% &
  1 \\
Touch followed By Gesture &
  0 &
  0\% &
  \multicolumn{1}{c|}{0} &
  6 &
  18\% &
  \multicolumn{1}{c|}{3} &
  2 &
  6\% &
  \multicolumn{1}{c|}{1} &
  0 &
  0\% &
  0 \\
Interleaved Touch and Conversation &
  \cellcolor[HTML]{FFCE93}10 &
  \cellcolor[HTML]{FFCE93}30\% &
  \multicolumn{1}{c|}{\cellcolor[HTML]{FFCE93}5} &
  \cellcolor[HTML]{FFCE93}10 &
  \cellcolor[HTML]{FFCE93}30\% &
  \multicolumn{1}{c|}{\cellcolor[HTML]{FFCE93}7} &
  \cellcolor[HTML]{FFCE93}10 &
  \cellcolor[HTML]{FFCE93}30\% &
  \multicolumn{1}{c|}{\cellcolor[HTML]{FFCE93}7} &
  2 &
  9\% &
  2 \\
Interleaved Touch and Gesture &
  7 &
  21\% &
  \multicolumn{1}{c|}{3} &
  3 &
  9\% &
  \multicolumn{1}{c|}{2} &
  5 &
  15\% &
  \multicolumn{1}{c|}{4} &
  2 &
  10\% &
  1 \\
Interleaved Touch, Conv., and Gesture &
  0 &
  0\% &
  \multicolumn{1}{c|}{0} &
  1 &
  3\% &
  \multicolumn{1}{c|}{1} &
  1 &
  3\% &
  \multicolumn{1}{c|}{1} &
  0 &
  0\% &
  0 \\ \hline
\textbf{Total} &
  33 &
  100\% &
  \multicolumn{1}{l|}{} &
  33 &
  100\% &
  \multicolumn{1}{l|}{} &
  33 &
  100\% &
  \multicolumn{1}{l|}{} &
  21 &
  100\% &
  \multicolumn{1}{l|}{} \\ \hline
\end{tabular}
\end{table*}

We identified nine distinct modality interaction patterns that were used during exploration and analysis tasks based on the modalities -- Touch, Touch Gesture, or Conversation -- and the order in which these modalities were used (Table~\ref{tab:interactionPatterns}). We distinguished between \textit{directed touch}, where the participants purposefully explored a graphic, and \textit{undirected touch}, which was more haphazard, and appeared to lack a purpose. 
The choice of interaction pattern strongly depended on the type of task and prior tactile experience. The presentation format (line chart, bar chart, or isarithmic map) did not appear to influence the interaction pattern and choice of modality.

\subsubsection{The Impact of Task}
Table~\ref{tab:interactionPatternsUse} shows the patterns that participants used for each task.

\bipstart{Initial exploration}
The \Water\ line charts were explored on average for 3:16 min, the \Water\ bar charts for 1:19 min, and the \Fire\ isarithmic maps for 3:07 min. The exploration of the \Stock\ line chart was excluded from the results, as it was used for training.

All participants utilized touch in the initial exploration of the \Water\ and \Fire\ visualizations. About half (49\%) of the exploration tasks were completed using \textit{Directed Touch Only}, \ie\ initial information gathering of the visualizations was performed purely through touch. Two other patterns, \textit{Interleaved Directed Touch and Conversation} (30\%) and \textit{Interleaved Directed Touch and Gesture} (21\%), were also used. While most time was still spent exploring via touch, five participants asked the conversational agent questions while exploring. For example, they asked for a description of the graphic or the difference between a line chart and a bar chart, and used speech to obtain descriptions of data values or axis labels before returning to touch.

\bipstart{Identifying specific or extreme values}
Even though the answer could be easily obtained by directly asking the conversational agent for these values, the \textit{Conversation Only} and \textit{Conversation and Undirected Touch} patterns were infrequently used. Instead, 87\% of the specific-value tasks and 81\% of the extreme-values tasks were completed using touch with speech or gestures such as double taps: \textit{Interleaved Directed Touch and Conversation}, \textit{Directed Touch followed by Conversation}, and to a lesser extent \textit{Directed Touch followed by Gesture} and \textit{Interleaved Directed Touch and Gesture}.

When asked to identify specific values, \eg\ the most recent rainfall in \Water\ or the number of fire-weather days in \City\ in \Fire, participants {generally built an initial understanding through touch}, \eg\ counting bars or \textit{x}-axis markers to identify a data point and tracing it to the \textit{y}-axis, before confirming values using either speech or a gesture (\textit{Directed Touch followed by Conversation [24\%] or by Gesture [18\%]}). Across the three specific-values tasks, seven participants also chose to verify this output by undertaking further \textit{Directed Touch}, employing the \textit{Interleaved Directed Touch and Conversation} (30\%) pattern, \eg\ comparing the found value to surrounding data points, or re-tracing to corresponding \textit{y}-axis points.

These patterns were used in similar ways when identifying extremum values, \eg\ the highest water storage in \Water\ or areas of the \Fire\ isarithmic map that showed decreases in fire-weather days. Participants would often start with touch, by hovering and rubbing their hands in circular motions to identify extremum charted values, before asking the conversational agent for exact values. In many cases, participants would then trace between close extrema to compare them, undertaking subsequent spoken or gesture-based interactions to extract values.

\bipstart{Describing a trend}
All participants were able to describe the general trend of the \Water\ and \Fire\ visualizations. Just under half (43\%) of the trend tasks were completed using \textit{Directed Touch Only} to build up an understanding of the visualizations. This included tracing the \Water\ line chart or bar chart, identifying key extrema, and comparing values in a systematic way. For \Fire, this involved tracing the isarithmic map's perimeter and rubbing hands in circular motions over the isarithmic map to identify areas with change. One participant (P5) described the \Fire\ trend verbally, without any pattern or explicit interaction (touch or otherwise) altogether, instead relying on the knowledge they had built during earlier tasks. 

The multimodal pattern \textit{Conversation followed by Directed Touch} was also used, but to a lesser degree (19\%). Three participants asked the conversational agent to describe a trend, followed by touching the visualizations to compare and seemingly validate the agent's answer.

\subsubsection{The Impact of Tactile Experience}
\label{subsection:tactileexperience}
The participants' self-reported tactile experience appeared to influence which interaction patterns they used when completing tasks (Table~\ref{tab:tactileQAStrategies}). One participant (P5) reported limited exposure to tactile graphics, and two (P4 and P9) reported no exposure (Table~\ref{tab:participants}). P9 spoke of how a lack of experience made them hesitant to rely on touching the visualizations, \textit{``... it takes me a while, I suppose because I am not really used to tactile.''} These participants appeared to consume information through touch in a way that had less definition, and at times used touch in a way that was undirected (goalless) or unsuccessful. P9 described difficulty visualizing what they were touching -- \textit{``I can't grasp what this is telling me very well.''} At the same time, P4 confused the isarithmic map for a bar chart, and P5 experienced difficulty locating a prominent island featured on the isarithmic map. There were also cases where these participants touched the RTD mid-refresh, which caused confusion. 

Participants with low tactile experience ($n=3$) were the only group that relied on the \textit{Conversation Only} (19\%) and \textit{Conversation and Undirected Touch} (16\%) patterns in a significant way (Table~\ref{tab:tactileQAStrategies}). In these interactions, the low tactile experience participants were satisfied just by asking the conversational agent for answers.

Participants with medium ($n=3$) and high ($n=5$) tactile experience were more comfortable completing tasks using \textit{Directed Touch Only}, choosing this modality to complete roughly a quarter of the tasks, and using touch with conversation and/or gestures to complete the remaining tasks. The only noticeable difference between these two groups was that participants with medium tactile experience preferred to use the conversational agent over gestures, whereas the more experienced participants tended to switch between them.

\begin{table}[t]
\renewcommand{\arraystretch}{1}
\setlength{\tabcolsep}{2.25pt}
\small
\centering
\caption{QA strategy use classified by self-reported level of tactile experience (Low [Limited/None] = P4, P5, P9; Medium [Some] = P6, P7, P8; High [Substantial] = P1, P2, P3, P10, P11); 
the top two patterns employed by each group, based on tactile experience,  are highlighted.\label{tab:tactileQAStrategies}}
\begin{tabular}{|l|cc | cc | cc|}
\hline
\multicolumn{1}{|c|}{} &
  \multicolumn{6}{c|}{\textbf{Tactile Experience}} \\ \cline{2-7} 
\multicolumn{1}{|c|}{\multirow{-2}{*}{\textbf{Interaction Modality Pattern}}} &
  \multicolumn{2}{c|}{Low ($n=3$)} &
  \multicolumn{2}{c|}{Med. ($n=3$)} &
  \multicolumn{2}{c|}{High ($n=5$)}\\ 
 & \# times & \% & \# times & \% & \# times & \%\\
  \hline
Conversation Only &
\cellcolor[HTML]{FFCE93}6 & \cellcolor[HTML]{FFCE93}19\% & 0 & 0\% & 1 & 2\% \\
Conversation and Undirected Touch &
5 & 16\% & 1 & 3\% & 0 & 0\% \\
Conversation followed By Touch &
2 & 6\% & 2 & 6\% & 1 & 2\% \\
Touch Only &
5 & 16\% & \multicolumn{1}{c}{\cellcolor[HTML]{FFCE93}8} &  \multicolumn{1}{c|}{\cellcolor[HTML]{FFCE93}24\%} &
  {\cellcolor[HTML]{FFCE93}14} &  {\cellcolor[HTML]{FFCE93}26\%} \\
Touch followed By Conversation &
2 & 6\% & \multicolumn{1}{c}{7} &  \multicolumn{1}{c|}{21\%} &
  8 & 15\% \\
Touch followed By Gesture &
2 & 6\% & 0 & 0\% & 6 & 11\% \\
Interleaved Touch and Conversation & \cellcolor[HTML]{FFCE93}6 & \cellcolor[HTML]{FFCE93}19\% & \multicolumn{1}{c}{\cellcolor[HTML]{FFCE93}13} & \multicolumn{1}{c|}{\cellcolor[HTML]{FFCE93}39\%} & \cellcolor[HTML]{FFCE93}12 & \cellcolor[HTML]{FFCE93}22\% \\
Interleaved Touch and Gesture &
4 & 13\% & 1 & 3\% & \cellcolor[HTML]{FFCE93}12 & \cellcolor[HTML]{FFCE93}22\% \\ 
Interleaved Touch, Conv., and Gesture &
0 & 0\% & 1 & 3\% & 1 & 2\% \\ \hline
  {\bf Total} & 32 & 100\% & 33 & 100\% & 55 & 100\% \\ \hline
\end{tabular}
\end{table}

\subsection{Conversational Queries}
\label{subsection:conversationalQueries}
Participants asked Graphy's conversational agent a total of 202 queries ({$\mu = 18.5, \sigma = 15.9$}), across four categories: 
\begin{itemize}
\vspace*{1mm}
    \myitem \textit{Data Extraction} (102, 50\%) queries related to the retrieval of data values, \eg\ \textit{``what is the value [that I am touching]?''} or \textit{``what was the rainfall in 2006?''}

    \myitem \textit{Operations} (69, 34\%) queries to manipulate the data or visualizations, \eg\ \textit{``can you zoom this area [of the graph]?''} or \textit{``can you transform this data into a bar graph?''}

    \myitem \textit{Overview} (20, 10\%) queries to summarize the data, \eg\ \textit{``describe this chart''} or \textit{``what is the trend showing on this chart?''}

    \myitem \textit{Understanding Visualization} (11, 5\%) queries that provide context relating to the type, layout or encoding of a visualization, \eg\ \textit{``how many segments are in this graph?''} or \textit{``what is a bar graph?''}
\end{itemize}

\subsection{Semi-Structured Interview and Thematic Analysis}
\label{subsection:Interview}
We asked participants to indicate their level of agreement with statements about whether the combination of the RTD and the conversational agent would make them more comfortable completing basic data analysis tasks and complex tasks, and also whether this combination offers significant benefits over each format 
alone---conversational agent, RTD, or tactile graphic (5-point Likert scale). We ran binomial tests to determine whether the difference between the number of participants who agreed or strongly agreed and those who were neutral or below is statistically significant. We performed Holm-Bonferroni corrections for multiple comparisons~\cite{Holm1979}.

\begin{figure}[t]
  \includegraphics[alt={A bar chart representing participants' views on a 5-point Likert scale (from Strongly Disagree to Strongly Agree) regarding the benefits of the RTD plus the conversational agent compared to just a conversational agent alone, just the RTD alone, and a tactile graphic alone. The bar that presents a comparison with just a conversational agent shows that all 11 participants answered Strongly Agree. The bar that represents comparison with just the RTD shows 8 Strongly Agree, 1 Agree, and 2 Neutral. The bar that presents a comparison with just a tactile graphic shows 7 Strongly Agree, 3 Agree and 1 Neutral.},width=1.00\columnwidth]{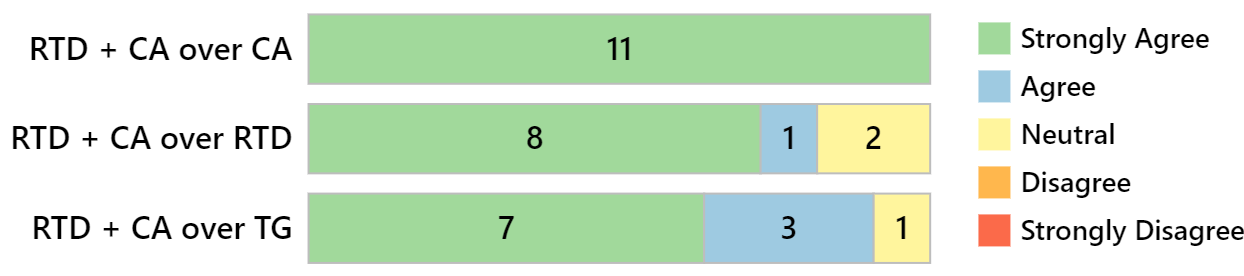}
  \caption{Participants' views on a 5-point Likert scale (1: Strongly Disagree and 5: Strongly Agree) regarding the benefits of the RTD plus the conversational agent (CA) compared to the CA alone, the RTD alone, and tactile graphics (TG) alone; the \textit{x}-axis represents number of participants.\label{fig:combinations}}
\end{figure}

Most participants agreed or strongly agreed with the statement about basic data analysis tasks ($k=8$, $n=11$, $p=0.029$), and more participants agreed or strongly agreed when asked about more complex tasks ($k=10$, $n=11$, $p=0.003$). In addition, participants showed a strong preference for the combination over just a conversational agent ($k=11$, $n=11$, $p<0.001$), RTD ($k=9$, $n=11$, $p=0.012$), or tactile graphic alone ($k=10$, $n=11$, $p=0.003$) (Figure \ref{fig:combinations}).

A thematic analysis of participant responses during the WOz study and interview provided further insights, yielding the following themes: \textit{usefulness and use cases, multimodality, deeper experiences through tactile visualization, autonomy, user choice}, and \textit{design feedback}. 

\bipstart{Usefulness and Use Cases} All 11 participants described ways they could integrate a device that combines an RTD and a conversational agent into their everyday lives, both work and personal.

P8 described how it would help \textit{``reduce the fear and anxiety of [working with] graphs and charts,''} suggesting they would like to use Graphy to access the railway heat maps they encounter at work. P1, who works as a data analyst, noted that the device would \textit{``make their life a whole lot easier,''} and suggested that it could be used to explore stock prices and perform technical analyses. P11 indicated that Graphy could display many of the charts and tables used in payroll duties.

Five participants shared stories of being under-equipped to participate in data literacy activities during schooling, and six suggested that access to a device combining an RTD and a conversational agent could have expanded tertiary opportunities, including statistics (P4, P1, P8), mathematics (P5), computing algorithms (P2), and psychology (P10).

Seven participants spoke of how Graphy could improve access to data for personal uses. This included personal finance and budgeting (P3, P7), monitoring and interpreting health data, such as blood sugar and blood pressure (P7), real-time sports data (P6), and weather forecasts (P9). P7, who became blind at 15, described the utility of Graphy well, \textit{``This is great. I would love a real world application of this. There is a lot of data I don't have access to that I did say 10 years ago.''}

\bipstart{Multimodality}
Four participants described how Graphy's multimodal nature was complementary. P8 spoke of how both the RTD and conversational agent were integral to their understanding of the visualizations, stating that \textit{``the RTD is tangible and the conversational agent provides the conceptual information. Touch is the strongest sense we have after sight to observe the physical world. The conversational agent then helps us to understand. The combination of both incorporates knowledge and experiment.''} P3 described the multimodal combination as having \textit{``so many advantages because it is more of a 360 degree view of the information.''} 

P7 discussed how the conversational agent allowed them to \textit{``connect the dots between what [they were] touching,''} while P6 preferred the combination that Graphy offered compared to printed tactile graphics, indicating that \textit{``the conversational agent allows me to get answers to my questions or have the confusion cleared up, while the printed page will remain silent and force me to guess [or move on].''}

\bipstart{Deeper Experiences through Tactile Visualization}
Several participants described how tactile graphics helped facilitate deeper experiences with data and allowed for independent interpretation. 

The \Fire\ isarithmic map, in particular, helped participants build a `visual image' in their head. P8 felt that after the session, the map would \textit{``be stuck in my mind, I can still visualize [the state of] Queensland and Australia,''} and continued, \textit{``my brain is [so] stimulated at the moment.''} P8 later discussed how the data would have less meaning to them if it were presented in another format, like a table. P1 became visibly excited, while P7 expressed emotion touching the isarithmic map, describing \textit{``I love the shape of Australia $\ldots$ that is something you really miss when you cannot see. Simple things like a map.''}

Three participants (P1, P2 and P6), all of whom had higher tactile experience, indicated how the tactile graphics supported independent interpretation. They did not want to simply \textit{``be told the answer,''} but instead preferred to build their own understanding of the data first \textit{through touch}. P1 repeatedly acknowledged that they could just ask Graphy for results, but first wanted to explore the visualizations and work out their own answers. P6 described their process, \textit{``my first instinct is to use my hands to first feel what the graph is doing... I would like to find the answer myself, then use voice to verify it, you feel more accomplished if you find [it] yourself.''} P2 even became visibly frustrated when Graphy gave too much information for their question, feeling that their independence of interpretation had not been respected -- \textit{``I didn't ask for an analysis... I just wanted to know what the data was... that would be the type of analysis I would [like to] do myself.''}

\bipstart{Autonomy}
Four participants mentioned how they valued that Graphy could provide independent access to data without the need for assistance. 
P8 described how \textit{``Graphy explains things better, otherwise [I would] require someone else who [can explain the concept],''} while P3 felt that the device had allowed them to build an understanding equivalent to what a helper would, but on their own. P6 felt that Graphy made them more autonomous, \textit{``it feels more independent... I don't need somebody to tell me the answer.''} P5 spoke of how in high school they had a support worker who would create sine graphs out of modelling clay, and suggested that the combination of RTD and a conversational agent could serve a similar purpose.

\bipstart{User Choice} Participants mentioned that the ways they would choose to use Graphy depended on purpose, personal preference, environment, and prior experience with technology.

Three participants discussed how their use would depend on the purpose of the data analysis. P1 described that it would \textit{``depend on what I am doing... how deeply I want to explore the data,''} stating that in an educational or work context they would engage in deeper interactions to ensure they completely understood what was graphed. P2 discussed varying interactions based on the effort required, \textit{``I am more likely to use the conversational agent for more complex tasks, [those where] the effort to benefit ratio is better,''} while P8 stated that for basic tasks they would rely on touch, but for work tasks would use touch gestures and voice interaction.

Four participants mentioned that the way they interacted was a  personal choice. P4 mentioned that \textit{``these are my needs, other people might think and interact differently,''} while P5 stated that \textit{``the [methods] are all useful, but I prefer buttons and touch.''} P3 connected their preference to level of ability, indicating that \textit{``because of paralysis, my hand can get sore, so designing it so I can use voice interaction is really important,''} while P8 favored voice interaction as \textit{``it is what I am most comfortable with.''}

Unsurprisingly, all participants described how environment might impact modality choice. P8 and P10 indicated that they would be hesitant to talk to Graphy while working in close proximity with co-workers, and suggested the need for headphones in these scenarios, while P6 said that in loud environments, they would \textit{``type commands, but at home [I] would talk to it.''} 

Prior experience with technology also played a role. This was strongest when it came to willingness to use touch gestures and Graphy's conversational agent. Participants with extensive experience with smartphones appeared more willing to incorporate touch gestures. P6, who chose to zoom into the visualizations using a double tap and hold, discussed how their choice \textit{``corresponded with iPhones.''} On the other hand, P8 described preferring interacting using voice, \textit{``I think I am so accustomed to my smart speakers and getting information that way.''} Two participants who did not use touch gestures often, but were comfortable relying on conversation during interactions, stated, \textit{``I am used to Google Assistant''} (P4) and \textit{``it is easiest to get the information I want... [by] asking a question''} (P9).

\bipstart{Design Feedback}
Participant feedback centered on labelling and complexity limits.
Seven participants desired more extensive braille labelling. While braille titles and legends were provided, participants wanted the axes of the visualizations to be labelled. Six of them stated that they should be able to directly read braille labels or tap on them to have Graphy read the labels out aloud. Four participants (P1, P2, P6, and P11) acknowledged that due to the space and resolution constraints of the RTD, it may be difficult to sufficiently label the visualizations, with P2 suggesting that a touch gesture could summon/hide labels, \textit{``displaying axis markers takes up real estate, so having those not showing all the time, being able to bring them up easily... a press gesture... would be useful.''} Another possibility, suggested by P2, was to render touch discernible symbols on the RTDs display, which users can tap to extract auditory labels.

Four participants (P1, P2, P6, and P8) felt that there was a limit on the complexity of visualizations that can be effectively understood on Graphy's RTD. This related to the number of pins and pin heights that the RTD currently supports. Prior to being introduced to the \Water\ line chart, participants were asked if they wanted to display water catchment storage, rainfall, or both data together. Only two participants (P1 and P3) chose to show them together when completing the analysis tasks, while three participants (P5, P8, and P10) speculated that displaying both datasets would prove overwhelming. Indeed, when asked to perform the adding data set operation, P8 reported difficulty interpreting sections of the multi-line chart where rainfall and water storage overlapped. 

P8 discussed how the RTD's pin heights were easy to read in isolation, \eg\ in a six dot braille cell, but difficult to read in close proximity to one another, and P6 said that in close proximity, the pin heights did not have enough \textit{`contrast.'} P1 and P2 focused on the low pin resolution of the RTD, describing the isarithmic map as a \textit{`pixelated version'} of Australia. P2 also felt that Graphy was capable of rendering visualizations at lower resolutions than what they were used to with tactile graphics.

Keyboard input was also requested by three participants, to allow shortcuts, or as an alternative to speech input and touch gestures when manipulating visualizations, \eg\ filtering or transforming. 

\section{Discussion and Limitations}
\label{section:discussion}
\subsection{The Combination of Conversational Agents and RTDs is Beneficial and Preferred}
\label{subsection:multimodalaccess}

Five participants shared stories of frustration when being under-equipped to participate in data literacy activities during school years, highlighting the need for accessible data tools. This aligns with prior research showing that BLV students are underrepresented in STEM disciplines in tertiary education, due to unequal access~\cite{butler2017understanding}. 

Participants' overall reaction toward the combination of RTD and conversational agent was very positive. They indicated that the combination would support both basic and more advanced data analysis. We observed that all participants used a mixture of touch,  touch gestures, and natural language queries in their interactions with data and visualization. There was also a clear preference for the combination of RTD and conversational agent over a conversational agent, RTD, or printed tactile graphic alone. 

To some extent, this preference for the combination of RTD and conversational agent is to be expected, given that multiple modalities increase adaptability~\cite{Reeves2004}. In line with this, our participants indicated that different modalities would allow them to cater to personal preferences and the task context and purpose. 

However, our results also revealed that prior experience with tactile graphics played a key role in the choice of modality.
In particular, participants with lower tactile graphic experience leveraged the conversational agent to a greater degree. This aligns with earlier research by Reinders \etal~\cite{Reinders2020}, where participants with lower tactile experience used a conversational interface embedded in a multimodal interactive 3D printed model earlier and more frequently than participants with higher experience. Nonetheless, all participants, even those with little tactile graphic experience, strongly preferred the combination of RTD and conversational agent over a conversational agent alone. Furthermore, one participant (P8) stated that over time, they may be more willing to use touch more frequently and confidently, suggesting that less experienced BLV users may build tactile confidence and adjust their choice of interaction after using the system. 

We were surprised by the strength of preference for tactile visualizations by users with more tactile experience. Participants spoke about how a tactile graphic allowed them to visualize the data, and most of those with greater tactile experience did not want to simply \textit{``be told the answer,''} preferring to assemble their own understanding of the data through touch first. This desire for independent understanding came in spite of the additional time required for tactile exploration. Independence of interpretation has been recognized as a critical factor for increasing confidence in data-based decision-making by BLV people~\cite{Fan2023}. For experienced tactile graphics users, the RTD provided BLV users with the opportunity for individual sense-making, just as visual graphics do for sighted users, in a manner that is complementary to information provided by a conversational agent. 

\subsection{Is Graphy Practical?}
A key question raised by any WOz study is whether the interface desired by participants can be constructed using current technologies. Because we catered for a broad range of queries pertaining to summary and background information about the visualizations (informed by~\cite{Kim2023}), almost all participant queries could be fulfilled using the scripts we prepared. The generation of these scripts was straightforward, and we believe could be readily supported by existing conversational agent platforms such as Dialogflow\footnote{https://cloud.google.com/dialogflow} and Rasa\footnote{https://rasa.com/docs/rasa/}.

One exception was the \Fire\ isarithmic map, when participants were asked to determine the number of fire-weather days for a specific city. One participant (P10) asked for the location of the city and was given directions generated by the wizard on the fly. Another instance occurred when P10 was exploring the map and asked the agent if there had been a change in dangerous fire weather in a small regional city they had once visited while holidaying. Answering questions such as this requires the conversational agent to have an understanding of background information about the data. In principle, this may be supported by means of large language models (LLMs), but it raises issues of reliability and trust. 

Participants also desired control over the flow of speech output, including the ability to pause, resume, or stop the conversational output. Placing users in control during conversational interactions has been identified as a critical need with mainstream conversational agents~\cite{Abdolrahmani2018,Branham2019}. Based on our experience designing conversational agents with BLV co-designers~\cite{Reinders2023}, this would be straightforward to support using Speech Synthesis Markup Language\footnote{https://www.w3.org/TR/speech-synthesis11/}.

Our study also suggested that participants valued the use of pin height and single-touch and multi-touch gestures. Unfortunately, cheaper RTDs, such as the Dot Pad, do not support variable-height pins, and are not touch-sensitive. Even touch-sensitive RTDs, such as the Graphiti or Monarch, only support single-finger touch gestures and require the user to lift their other fingers when completing gestures. Thus, how best to support touch gestures remains an open issue.

\subsection{How Do We Design Interactive Tactile Visualizations?}
One area that needs further research before we can build an actual system, is a better understanding of how to design interactive tactile visualizations for an RTD. While there exists well-established guidance on visualization best practices when it comes to different types of data and tasks~\cite{Munzner2015}, such guidelines do not exist for the design of tactile visualizations for RTDs. 

For instance, at present, we do not know the relative effectiveness of tactile line graphs and bar charts for different tasks, or the affordances they offer in this context. In our study, we observed that participants tended to prefer bar charts over line graphs. P6 described \textit{``the bar chart is easier to follow, I think it [allows me to] more accurately feel data at any particular point.''} This seemed to be particularly true for participants with lower tactile experience, with P9 stating that \textit{``[with] the bar chart [it] was a lot easier to determine changes, [both the] highs and lows.''} We observed that participants tended to trace line graphs using one or two fingers, with many tracing back and forth between inflection points, and that they traced up and down individual bars in a bar chart, and that in some cases, two hands were used to trace multiple bars. This hints that, just as for visual line graphs and bar charts~\cite{Munzner2015}, tactile line charts are suitable for exploration of trends, while bar charts are suitable for discrete comparisons. However, more research is required to test this.

While there are guidelines for the design of traditional tactile graphics (\eg~\cite{BANA2010guidelines}), there are significant differences between these and tactile graphics displayed on RTDs. One important difference is the current low resolution of RTDs, and the consequent inability to use braille labels, \eg\ markers and values on axes. Participants identified possible solutions, but more research is required.

Another area requiring research is how best to support zooming and panning. Touch gestures were used by six participants to pan (swipe) and zoom (pinch). However, when panning or zooming, five participants (P1, P2, P6, P7, and P8) became confused and lost their location when the RTD was refreshed. P1 said \textit{``Wow... that is crazy... I lost where I was.''} P7 spoke of how \textit{``when zooming, [you need] to not lose your frame of reference, to be able to find it again.''} P2 directly suggested that a scroll bar could be used to avoid this, \textit{``if you are panning a graph you could have an indicator where spatially in the graph you are, for example, you could have one column and row of pins, if you're viewing the bottom then the column would be lower. Like a scroll bar would show where you are in a page.''} Previous research with RTDs has suggested the use of buttons, rather than gestures, for operations like zooming to avoid context loss and user confusion~\cite{zeng2014examples,Zeng2015}. More research is needed into ways in which multimodal feedback can be used to retain user understanding of position and context during interactions such as panning and zooming, filtering, and search.

A further interaction warranting investigation is using pin actuation to highlight points or regions of interest on the RTD. Actuation has previously been used to highlight routes on maps~\cite{Holloway2022}. We did not support actuation in our study but actuation could, for instance, support search and multimodal output where data points referenced in audio output are concurrently actuated in the visualization. This would be analogous to the use of speech with haptic vibratory feedback in interactive 3D models~\cite{Reinders2023}. Such multimodal output would have been particularly useful with the \Fire\ isarithmic map, when a participant asked for the location of a city and was given directions from the agent, which would have been more effective if combined with actuation. 

\subsection{Limitations}
As the first step to understand the benefits of the combination of an RTD and conversational agent, our study involved only three visualization types -- line graphs, bar charts, and isarithmic maps -- and four task types -- exploration, value identification, extremum identification, and trend. Future research needs to consider a broader range of visualization types and examine additional tasks users perform with these visualizations. 
We also primarily focused on the presentation of univariate spatial and time series data. Visualizations of more complex multivariate data, \eg\ multiple time series visualizations, should be investigated.

As the majority of our participants were totally blind, we are also interested in recruiting additional legally blind participants, specifically those who still have residual vision, as their perspective may differ from that of people who are totally blind.

Our findings should also be interpreted within the limitations of the WOz method. While our findings strongly suggest that a combination of RTD and conversational agent is an effective tool for data analysis, it is crucial to implement and evaluate an actual system with BLV users. It would also be important to test the system with participants over a longer period of time and in real-world settings, \eg\ in-the-wild studies in schools and workplaces.

\section{Conclusion}
Supporting BLV people's data analysis remains an important and challenging question. Here, we investigated a novel approach that combines an RTD with a conversational agent by conducting a Wizard-of-Oz study with 11 BLV participants. 

From the analysis of participants’ interactions, we identified a variety of multimodal interaction patterns. They depended on tasks and participants' familiarity with tactile graphics. However, participants also made it clear that other factors such as purpose, location, and prior experience with smart phones and conversational agents influenced their interactions.

Our results show that all participants blended different modalities. Furthermore, participants suggested that the RTD and agent are complementary, and strongly believed that the multimodal combination supported autonomous data analysis and offered significant benefits over an RTD-only or a conversational-agent-only interface, or traditional tactile graphics.

We believe that the observations and findings, as well as participant suggestions from our study provide a solid basis for the design of a system supporting a combination of RTD and conversational agent, one that can allow BLV people to meaningfully engage independently in data access and analysis. 


\acknowledgments{We gratefully acknowledge the Australian Research Council's Discovery Projects funding scheme (DP220101221). This work was supported in part by the Institute of Information and Communications Technology Planning and Evaluation (IITP) Grant funded by the Korean Government (MSIT), Artificial Intelligence Graduate School Program, Yonsei University, under Grant RS-2020-II201361.}

\bibliographystyle{abbrv-doi-hyperref}

\bibliography{references}
\end{document}